\documentclass[notitlepage,showkeys,twocolumn,11pt,times,superscriptaddress, a4paper, pre, nofootinbib, 11pt]{revtex4-2}
\usepackage[english]{babel}
\usepackage{stix}
\usepackage[T1]{fontenc}
\usepackage{graphicx}
\usepackage{bm} 
\usepackage[colorlinks=true,
            linkcolor=blue,
            urlcolor=blue,
            citecolor=blue]{hyperref}
\usepackage[mathlines]{lineno} 
\usepackage{amsmath}
\usepackage{physics}
\usepackage{booktabs}
\usepackage{geometry}
\newcommand*\dif{\mathop{}\mathrm{d}}

\begin{document}

    \title{Hierarchical fragmentation of regular islands in a discontinuous nontwist map}
    \author{Matheus Rolim Sales}
    \email{rolim.sales.m@gmail.com}
    \affiliation{São Paulo State University (UNESP), Institute of Geosciences and Exact Sciences, 13506-900, Rio Claro, SP, Brazil}

    \author{Michele Mugnaine}
    \affiliation{Lorena School of Engineering (EEL-USP), University of São Paulo, 12602-810, Lorena, SP, Brazil}
    
    \author{Leonardo Costa de Souza}
    \affiliation{Department of Physics, Institute for Complex Systems and Mathematical Biology, SUPA, University of Aberdeen, AB24 3UX, Aberdeen, United Kingdom}
    \affiliation{Institute of Physics, University of São Paulo, 05315-970, São Paulo, SP, Brazil}

    \author{Iberê Luiz Caldas}
    \affiliation{Institute of Physics, University of São Paulo, 05315-970, São Paulo, SP, Brazil}
    
    \author{Edson Denis Leonel}
    \affiliation{São Paulo State University (UNESP), Institute of Geosciences and Exact Sciences, 13506-900, Rio Claro, SP, Brazil}
    
    \author{José Danilo Szezech Jr.}
    \affiliation{Graduate Program in Science, State University of Ponta Grossa, 84030-900, Ponta Grossa, PR, Brazil}
    \affiliation{Department of Mathematics and Statistics, State University of Ponta Grossa, 84030-900, Ponta Grossa, PR, Brazil}
    \date{\today}

    \begin{abstract}
        
    The destruction of regular regions in two-dimensional, area-preserving maps is traditionally described in terms of the breakup of invariant curves and the persistence of transport barriers. Here, we investigate how this scenario changes when continuity is lost. We study the extended standard nontwist map with a perturbation whose period differs from a full revolution on the cylinder. In this setting, the induced map on the cylinder becomes discontinuous, even though the map remains smooth on the real line. Using complementary chaos diagnostics, we find that regular islands are not enclosed by a single invariant curve but instead undergo hierarchical fragmentation into smaller regular components connected by chaotic channels. We show that trajectories initialized near elliptic points exhibit long trapping followed by escape, ruling out the existence of a global transport barrier. The fragmentation occurs when island chains are centered on the discontinuity line, while island chains away from it preserve the conventional islands-around-islands structure. By restoring continuity of the induced map on the cylinder in a modified formulation, we recover smooth invariant curves and eliminate fragmentation, demonstrating that the hierarchical structure originates from discontinuity rather than twist violation alone. Similar behavior is also observed in other two-dimensional area-preserving maps, indicating that the phenomenon is not restricted to nontwist systems.
    \end{abstract}

    \maketitle


    \section{Introduction}
    \label{sec:introduction}
    
    A two-degrees-of-freedom Hamiltonian system can be reduced to a two-dimensional, area-preserving map by using energy conservation and introducing an appropriate Poincaré surface of section~\cite{lichtenberg2013regular}. In general, the phase space of the resulting map is characterized by the coexistence of integrable and hyperbolic regions. Then, it exhibits a mixed structure, in which regular regions organized around invariant curves coexist with chaotic regions. In two-dimensional, area-preserving maps, invariant curves play a central role in shaping transport. They act as absolute barriers, preventing chaotic trajectories from crossing between distinct chaotic regions of phase space. The destruction of these invariant curves under perturbations is classically described by the Kolmogorov-Arnold-Moser (KAM) and Poincaré-Birkhoff theorems~\cite{lichtenberg2013regular}. As the perturbation strength increases, invariant curves break, resonant island chains appear, and chaotic regions expand. Consequently, the global transport of chaotic trajectories is directly related to the organization and persistence of regular structures within the chaotic sea.
    
    Even after invariant curves are destroyed, their remnants continue to influence transport. The cantori~\cite{Mackay1984, Mackay1984b, Efthymiopoulos1997}, which arise from the breakup of invariant curves, act as partial barriers that slow down transport without completely blocking it. This interplay between regular regions, invariant structures, and chaotic motion gives rise to the phenomenon of stickiness~\cite{Efthymiopoulos1997, Contopoulos1971, Meiss1983, Karney1983, Chirikov1984, Meiss1985, Meiss1986, Afraimovich1997, Zaslavsky2002, CONTOPOULOS2008, Cristadoro2008, Venegeroles2009, Contopoulos2010}. Chaotic trajectories that approach regular regions may remain trapped in their vicinity for long times, during which their dynamics closely resembles quasi-periodic motion, before eventually returning to the chaotic sea.
        
    The existence of cantori has been rigorously established for nondegenerate Hamiltonian systems, in which the frequency of motion depends monotonically on the action variables~\cite{lichtenberg2013regular}. For two-dimensional maps, this nondegeneracy requirement is expressed by the twist condition, defined as the nonvanishing of the shear, $\partial x_{n+1} / \partial y_n \neq 0$, throughout phase space. Maps that satisfy this condition everywhere are referred to as \textit{twist} maps, while those that violate it at one or more locations are classified as \textit{nontwist} maps. Violations of the twist condition introduce qualitatively new dynamical features, including shearless invariant curves and reconnection phenomena, which have no counterpart in twist systems. A paradigmatic example is the standard nontwist map (SNM)~\cite{del-Castillo-Negrete1993, del-Castillo-Negrete1996}, which has become a canonical model for studying the dynamics of nontwist systems and the universal properties associated with twist violation. Although some analytical progress has been made using extensions of KAM theory~\cite{Delshams2000}, most results for nontwist maps rely on numerical simulations~\cite{Howard1984, delcastillonegrete1997, HOWARD1995256, Shinohara1997, Corso1997, Morrison2000}, especially transport-related properties~\cite{Szezech2009, Szezech2012, Mugnaine2018, Mugnaine2020, LCSouza2023, Mugnaine2024, RolimSales2024, Baroni2024, Baroni2025, RolimSales2025}.
        
    Despite the additional dynamical features introduced by twist condition violation, the vast majority of studies on nontwist maps assume that the induced map on the cylinder is smooth and well defined. In these systems, the forcing terms are periodic in the angular variable, so the induced dynamics defines a smooth, area-preserving transformation on $\mathbb{T}\times\mathbb{R}$. Under these assumptions, the destruction of regular regions follows the usual scenario: invariant curves break through resonances, island chains form, and transport is mediated by cantori and chaotic layers, even though the detailed mechanisms may differ from the twist case. Much less is known when the induced map on the cylinder loses continuity. Area-preserving maps with explicit discontinuities or piecewise smooth definitions have nevertheless been investigated in several contexts, including sawtooth maps~\cite{Bird1988, Ashwin1997}, piecewise linear standard maps~\cite{Bullett1986}, and piecewise parabolic Frenkel--Kontorova models~\cite{Aubry1983}. In these systems, the loss of smoothness or continuity is introduced explicitly through a piecewise definition of the map or of the underlying potential. Here, instead, the map remains infinitely differentiable on $\mathbb{R}$, but the induced map on the cylinder loses continuity after the modulo projection because the perturbation is not compatible with the $1$-periodic identification of the angular variable. Therefore, the mechanism considered in this work differs fundamentally from the usual nonsmooth or piecewise-defined maps studied in the literature: the map is locally smooth everywhere on $\mathbb{R}$, but the induced map on $\mathbb{T}\times\mathbb{R}$ becomes discontinuous because the perturbation is not $1$-periodic in the angular variable. In this situation, the classical framework based on invariant curves, cantori, and enclosing transport barriers may no longer apply. In particular, it is not clear whether regular islands can still be bounded by a single outer invariant curve, or whether their destruction proceeds through a different mechanism altogether.
        
    Therefore, motivated by these questions, in this paper we investigate how the loss of continuity of the induced map on the cylinder affects the structure of regular regions in a paradigmatic nontwist system. We focus on the extended standard nontwist map (ESNM)~\cite{Portela2007, Wurm2013}, which is an extension of the SNM through an additional perturbation term whose periodicity depends on a parameter $m$. When $m$ is noninteger, the perturbation remains smooth on the real line but is no longer periodic with unit period, so the induced dynamics on the cylinder becomes discontinuous. In a previous work~\cite{RolimSales2025}, we showed that this discontinuity produces highly nontrivial transport properties and unusual manifold configurations. Here, we investigate the phase-space structures underlying these phenomena. By varying the perturbation strength and the parameter $m$, we analyze how regular islands are modified, whether transport barriers survive, and how the destruction of regular regions differs from the standard scenario of smooth area-preserving maps. We show that island chains centered at the discontinuity line undergo a fragmentation process in which the regular region is no longer enclosed by a single invariant curve, but instead becomes organized into disconnected smaller structures separated by chaotic channels. To demonstrate that this phenomenon is not restricted to the ESNM, in the Supplementary Material we present additional examples in other area-preserving maps whose induced dynamics on the cylinder is discontinuous.
        
    This paper is organized as follows: In Sec.~\ref{sec:model}, we introduce the ESNM and present its phase space structure for noninteger values of $m$. In Sec.~\ref{sec:dynamicalcharac}, we use complementary chaos indicators, the escape times, the largest Lyapunov exponent~\cite{Benettin1980, Wolf1985, Eckmann1985}, and the smaller alignment index (SALI)~\cite{Skokos2003, Skokos2004}, to characterize the dynamical properties of the islands. Section~\ref{sec:ftrte} focuses on the intermittent dynamics inside the islands, analyzed through a finite-time recurrence time entropy (RTE) approach~\cite{Sales2023}. Section~\ref{sec:why} discusses the mechanism responsible for the fragmentation and presents additional examples illustrating the generality of the phenomenon. In Sec.~\ref{sec:largercylinder}, we introduce a modified version of the ESNM defined on a larger cylinder, where the induced dynamics on the enlarged cylinder becomes continuous for rational values $m = p / q$, providing a continuous nontwist reference system. Finally, Sec.~\ref{sec:concl} summarizes our results and presents the conclusions.
    
    \section{The model}
    \label{sec:model}
    
    The Hamiltonian function of a quasi-integrable, autonomous system with two degrees of freedom can be written, in terms of action-angle variables $(\mathbf{y},\mathbf{x}) \equiv (y_1,y_2,x_1,x_2)$, as~\cite{lichtenberg2013regular}
    \begin{equation}
        \label{eq:Hamiltonian}
        H(y_1,y_2,x_1,x_2)
        =
        H_0(y_1,y_2)
        +
        \varepsilon H_1(y_1,y_2,x_1,x_2),
    \end{equation}
    where $H_0$ is an integrable Hamiltonian, $H_1$ represents a perturbation, and $\varepsilon$ controls the perturbation strength. The trajectories generated by Eq.~\eqref{eq:Hamiltonian} evolve in a four-dimensional phase space. Since the system is autonomous and, assuming a kinetic energy quadratic in the velocities, the Hamiltonian coincides with the total mechanical energy, $H=T+V\equiv E$, which is conserved, $\dif{H}/\dif t=\partial H/\partial t=0$. Energy conservation reduces the effective dimensionality by one: one variable (e.g., $y_2$) can be expressed in terms of the remaining three and the fixed energy $E$, so that trajectories with energy $E$ are confined to a three-dimensional energy surface embedded in the four-dimensional phase space. A further reduction is obtained by introducing a suitable Poincaré surface of section (PSS). For a broad class of quasi-integrable Hamiltonian systems, choosing the section defined by $x_2=\mathrm{const}$ and projecting onto the $(x_1,y_1)$ plane yields a two-dimensional map $\mathcal{M}:\mathbb{T}\times\mathbb{R}\to\mathbb{T}\times\mathbb{R}$, where $\mathbb{T} = \mathbb{R}/\mathbb{Z}$, of the form~\cite{lichtenberg2013regular}
    \begin{equation}
        \label{eq:generalmapping}
        \begin{aligned}
            y_{n+1} &= y_n + \varepsilon f(x_n,y_{n+1}),\\
            x_{n+1} &= x_n + \Omega(y_{n+1}) + \varepsilon g(x_n,y_{n+1}) \pmod{1},
        \end{aligned}
    \end{equation}
    where $f$ and $g$ are periodic functions of their arguments (with period $2\pi$, where $\mathbb{T} = \mathbb{R}/(2\pi\mathbb{Z})$ or $1$, where $\mathbb{T} = \mathbb{R}/\mathbb{Z}$), and $\Omega(y)$ is the rotation number. The derivative $\dif\Omega/\dif y$ defines the shear of the map. Equation~\eqref{eq:generalmapping} relates successive intersections of a trajectory with the PSS. The map is area preserving provided that $f$ and $g$ satisfy
    \begin{equation*}
        \frac{\partial f}{\partial y_{n+1}} + \frac{\partial g}{\partial x_n} = 0.
    \end{equation*}
    
    \begin{figure*}[t]
        \centering
        \includegraphics[width=\linewidth]{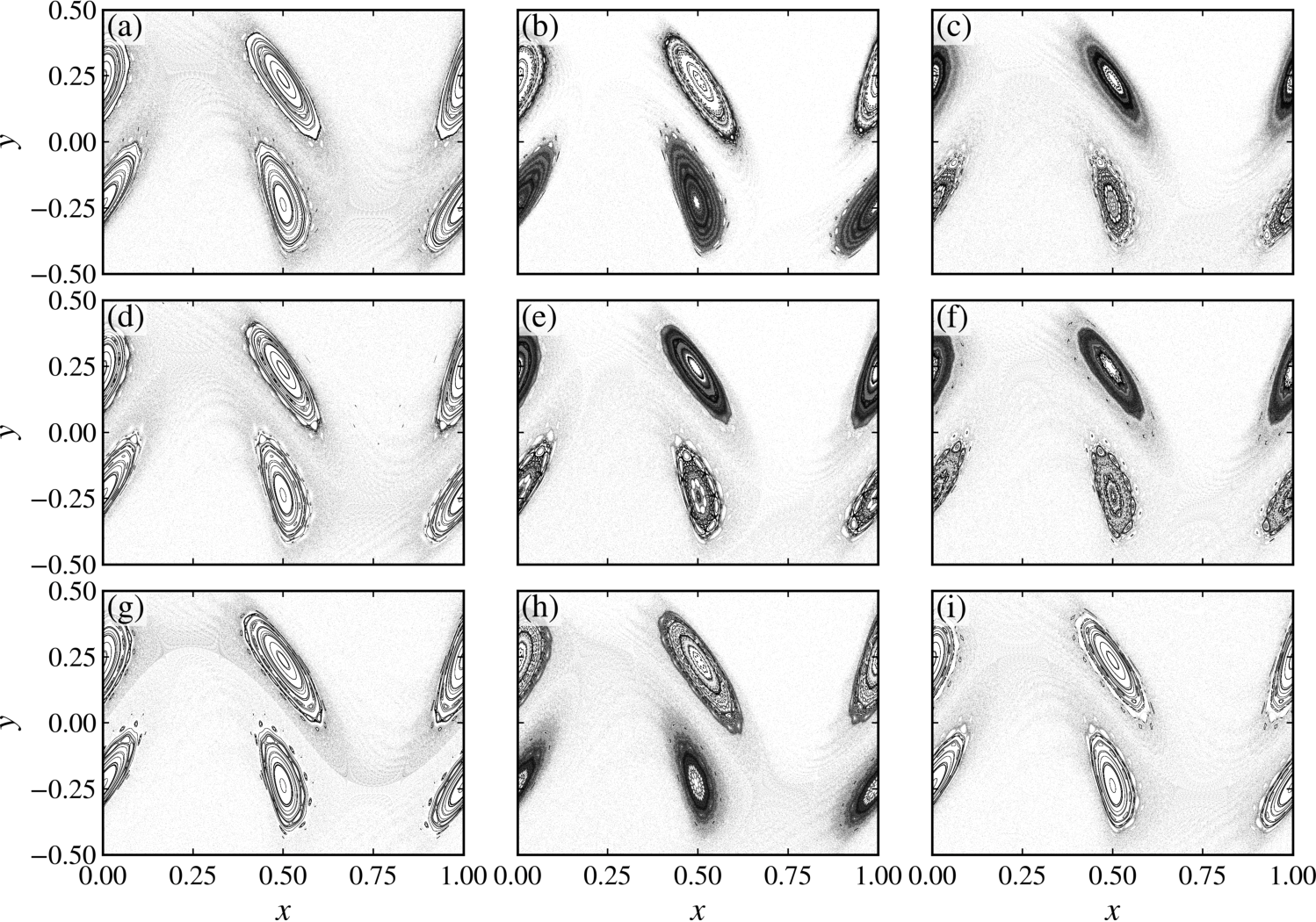}
        \caption{Phase space of the extended standard nontwist map [Eq.~\eqref{eq:esnm}] for $a = b = 0.53$, $c = 0.005$ and (a) $m = 0.0$, (b) $m = 0.4$, (c) $m = 0.8$, (d) $m = 1.0$, (e) $m = (\sqrt{5} + 1) / 2$, (f) $m = 1.8$, (g) $m = 2.0$, (h) $m = 2.3$, and (i) $m = 3.0$.}
        \label{fig:phase_space}
    \end{figure*}

    \begin{figure*}[t]
        \centering
        \includegraphics[width=\linewidth]{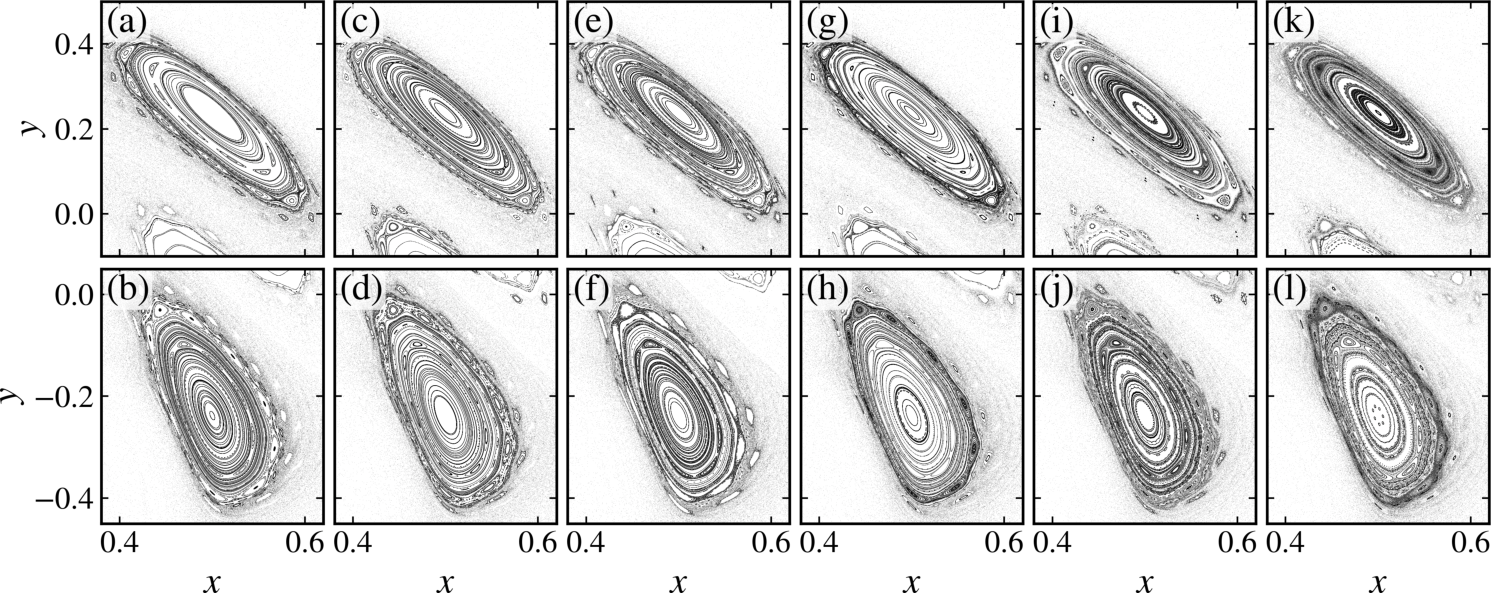}
        \caption{Magnifications around the upper (top row) and lower (bottom row) central islands for $a = b = 0.53$, $m = 0.8$, and (a) and (b) $c = 0$, (c) and (d) $c = 1.0\times10^{-5}$, (e) and (f) $c = 5.0\times10^{-5}$, (g) and (h) $c = 1.0\times10^{-4}$, (i) and (j) $c = 5.0\times10^{-4}$, and (k) and (l) $c = 1.0\times10^{-3}$.}
        \label{fig:phase_space_zoom}
    \end{figure*}
    
    A common and convenient choice is to set $g\equiv 0$ and take $f$ as a function of the angle variable only, for instance $f(x_n)=\sin x_n$, which automatically satisfies the area-preservation condition. Different choices of the rotation function $\Omega(y)$ lead to a variety of well-known models, including the standard map ($\Omega(y)=y$)~\cite{chirikovstdmap}, the static Fermi--Ulam model ($\Omega(y)=2/y$)~\cite{fermiulam1,fermiulam2}, the bouncer model ($\Omega(y)=\delta y$)~\cite{bouncermodel1,bouncermodel2}, the logistic twist map ($\Omega(y)=y+ay^2$)~\cite{HOWARD1995256}, the standard nontwist map ($\Omega(y)=a(1-y)^2$)~\cite{del-Castillo-Negrete1993,del-Castillo-Negrete1996}, and the Leonel map, a generalization of the Fermi--Ulam model ($\Omega(y)=1/|y|^{\gamma}$)~\cite{Leonel2009,deOliveira2010,deOliveira2013,Leonel2020,Borin2023}, among others. In this context, a map is said to be \emph{twist} if the rotation number is a monotonic function of the action variable, either strictly increasing or strictly decreasing. This requirement can be expressed as
    \begin{equation}
        \label{eq:twistcond}
        \left|
        \frac{\partial x_{n+1}}{\partial y_n}
        \right|
        =
        \left|
        \frac{\partial \Omega(y_{n+1})}{\partial y_n}
        \right|
        > 0,
    \end{equation}
    for all $(x_n,y_n)$. If the condition~\eqref{eq:twistcond} is violated in any region of phase space, the map is classified as \emph{nontwist}.
    
    In this work, we focus on the extended standard nontwist map (ESNM)~\cite{Portela2007,Wurm2013}, defined by
    \begin{equation}
        \label{eq:esnm}
        \begin{aligned}
            y_{n+1} &= y_n - b\sin(2\pi x_n) - c\sin(2\pi m x_n),\\
            x_{n+1} &= x_n + a\!\left(1-y_{n+1}^2\right) \pmod{1}.
        \end{aligned}
    \end{equation}
    Here $a$, $b$, $c$, and $m$ are real-valued parameters. The ESNM reduces to the standard nontwist map (SNM)~\cite{del-Castillo-Negrete1993,del-Castillo-Negrete1996} when either $c=0$ or $m=0$. In its original formulation, the parameter $m$ was restricted to positive integers~\cite{Portela2007,Wurm2013}. More recently, however, transport properties of the ESNM have been investigated for real values of $m$~\cite{RolimSales2025}, extending the results for the integer case~\cite{Mugnaine2020}. It is crucial to mention that even though the forcing term $f(x) = -b\sin(2\pi x) -c\sin(2\pi mx)$ is smooth on the real line $\mathbb{R}$, any noninteger $m$ makes the ESNM discontinuous on the cylinder $\mathbb{T}\times\mathbb{R}$, where $\mathbb{T} = \mathbb{R}/\mathbb{Z}$. Specifically, $f(x)$ is $q$-periodic on $\mathbb{R}$ for rational $m = p/q$ (in lowest terms), while irrational $m$ yields a quasi-periodic forcing on $\mathbb{R}$. In both cases, the forcing term is not $1$-periodic, so the map is not continuous after the modulo-$1$ identification that defines the cylinder.

    Therefore, for noninteger $m$, the ESNM is not a nonsmooth map in the usual sense of piecewise smooth dynamical systems. The forcing term remains infinitely differentiable on $\mathbb{R}$, and no local singularities or derivative discontinuities are introduced. The loss of continuity arises only after identifying points modulo $1$ in the angular variable. Consequently, the map is smooth on $\mathbb{R}$ but discontinuous on $\mathbb{T}\times\mathbb{R}$.
    
    Our goal is to investigate precisely how this discontinuity affects the dynamics of the ESNM. In all numerical simulations, we fix the parameters at $a = b = 0.53$. For the SNM ($c = 0$ or $m = 0$), these values give rise to two chains of distinct period-$2$ islands [Fig.~\ref{fig:phase_space}(a)]. We then vary $c$ and $m$ to introduce the discontinuity. Figure~\ref{fig:phase_space} shows the phase space of the ESNM for several rational (and one irrational) values of $m$, with $c = 0.005$. Panels (a), (d), (g), and (i), corresponding to integer values of $m$, represent the continuous case and are shown for reference. In all other cases (noninteger $m$), the islands exhibit clear qualitative differences from the continuous case. Some islands display dense accumulations of points around them [e.g., the lower island chain in Fig.~\ref{fig:phase_space}(b)], visible as darker regions associated with sticky chaotic motion, while others lose the smooth nested curves observed in the continuous case and instead develop more irregular internal structures [e.g., the upper island chain in Fig.~\ref{fig:phase_space}(e)]. 
    
    The detailed organization of these structures is investigated in Fig.~\ref{fig:phase_space_zoom}, which shows magnifications of the central (upper and lower) islands for $m = 0.8$ and different values of $c$. Panels (a) and (b) depict the continuous case ($c = 0$) from the upper and lower islands, respectively. As $c$ increases, the invariant curves gradually break down, giving way to small chaotic regions and chains of smaller islands, although some curves appear to persist even for $c > 0$. The novelty here is not the presence of darker regions or chains of smaller islands per se. Such features are typical in area-preserving maps and are known to lead to the stickiness effect. Usually, these island chains organize themselves into a hierarchical structure, with large islands surrounded by smaller ones, which are in turn surrounded by even smaller islands, and so on and so forth. Consequently, the darker regions (where trajectories become trapped) are generally located around these chains. In our case, however, we observe a diffuse blur of dark points occupying most of the area where the original ($c = 0$) island used to be. Moreover, while a hierarchical islands-around-islands structure persists, it manifests itself as a web-like tangle rather than the clean, self-similar structure seen in Figs.~\ref{fig:phase_space_zoom}(a) and \ref{fig:phase_space_zoom}(b) with $c=0$, for instance.
    
    \section{Dynamical characterization of the islands}
    \label{sec:dynamicalcharac}
    
    \begin{figure*}[t]
        \centering
        \includegraphics[width=\linewidth]{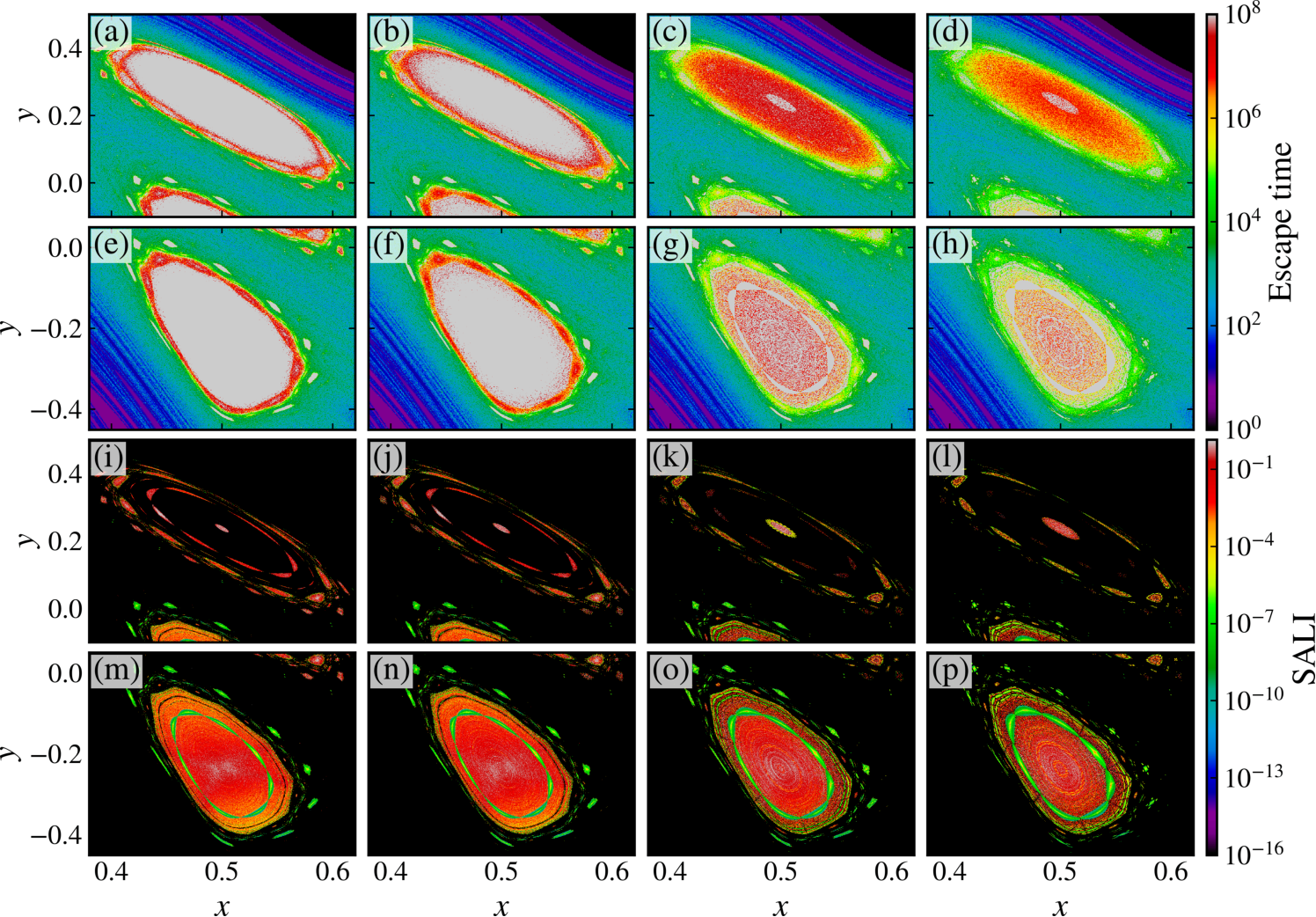}
        \caption{(a)--(h) Escape time and (i)--(p) the smaller alignment index (SALI) for a uniformly distributed grid of initial conditions around the (a)--(d) and (i)--(l) upper island and the (e)--(h) and (m)--(p) lower island for $a = b = 0.53$, $m = 0.8$, and (a, e, i, m) $c = 5.0\times10^{-5}$, (b, f, j, n) $c = 1.0\times10^{-4}$, (c, g, k, o) $c = 5.0\times10^{-4}$, and (d, h, l, p) $c = 1.0\times10^{-3}$. For the escape time, each initial condition is iterated (up to $10^8$ iterations) until it escapes the region defined by $y \in [-1, 1]$. The non-escaping initial conditions are colored as gray.}
        \label{fig:escape_time}
    \end{figure*}
    
    To quantitatively investigate the behavior described in Sec.~\ref{sec:model}, we compute the escape time for a $1000\times1000$ grid of initial conditions. Specifically, we consider trajectories originating within the regions shown in Fig.~\ref{fig:phase_space_zoom} and measure the number of iterations required for them to reach either $y = 1$ or $y = -1$, up to a maximum of $10^8$ iterations. The escape times are shown in Figs.~\ref{fig:escape_time}(a)--\ref{fig:escape_time}(h). In this representation, initial conditions that quickly escape are colored from black to blue, those with intermediate escape times range from green to yellow, and those with very long escape times appear in red. Initial conditions that do not escape within the maximum iteration time are shown in gray. For small $c$ [Figs.~\ref{fig:escape_time}(a) and \ref{fig:escape_time}(e)], almost no trajectory escapes from within the island up to $10^8$ iterations. As $c$ increases, more trajectories escape, although typically after extremely long times of the order of $10^7$. A naive interpretation would be that the non-escaping trajectories in Figs.~\ref{fig:escape_time}(a, b, e, f) correspond to regular motion, which is generally expected for continuous, two-dimensional, area-preserving maps. However, it is also possible that these trajectories are chaotic but require longer times to escape.
    
    To clarify this point, we compute the smaller alignment index (SALI)~\cite{Skokos2003, Skokos2004} for the same grid of initial conditions to distinguish between regular and chaotic dynamics. The computation of SALI proceeds as follows. Consider a $d$-dimensional discrete-time dynamical system $\Phi^n(\mathbf{x}_0)$ in $\mathbb{R}^d$, and let $\mathcal{D}\Phi^n(\mathbf{x}_0)$ denote the $n$th iterate of its Jacobian matrix. The deviation (tangent) vectors evolve according to
    \begin{equation}
        \label{eq:devvec}
        \mathbf{w}(n) = \mathcal{D}\Phi^n(\mathbf{x}_0)\mathbf{w}(0).
    \end{equation}
    The general solution of Eq.~\eqref{eq:devvec} can be written as
    \begin{equation*}
        \mathbf{w}(n) = \sum_{i = 1}^{d} c_i e^{\lambda_i n}\mathbf{\hat{e}}_i,
    \end{equation*}
    where $\lambda_i$ are the Lyapunov exponents and $\mathbf{\hat{e}}_i$ are the corresponding unit directions, where the hat notation means a unit vector: $\norm{\mathbf{\hat{e}}} = 1$. To compute SALI, we follow the evolution of two distinct deviation vectors along the orbit. SALI quantifies whether the corresponding normalized (unit) vectors become linearly dependent, that is, whether they tend to align, and is defined as
    \begin{equation}
        \mathrm{SALI}(n) = \min\left\{\left\|\mathbf{\hat{w}}_1 - \mathbf{\hat{w}}_2\right\|, \left\|\mathbf{\hat{w}}_1 + \mathbf{\hat{w}}_2\right\|\right\}.
    \end{equation}
    Geometrically, SALI measures the area of the parallelogram formed by the two vectors $\mathbf{\hat{w}}_1$ and $\mathbf{\hat{w}}_2$. Its behavior differs depending on whether the orbit is regular or chaotic. For two-dimensional, area-preserving maps, SALI decays algebraically for regular motion as $\mathrm{SALI} \propto n^{-p}$, with typically $p \approx 2$~\cite{Manos2008, Manos2008b}. In contrast, for chaotic orbits SALI decays exponentially in time according to the two largest Lyapunov exponents as $\mathrm{SALI} \propto e^{-(\lambda_1 - \lambda_2)n}$~\cite{Skokos2007, Sales2026}. Since SALI for chaotic orbits decays much faster, regular and chaotic dynamics can be distinguished by computing SALI for a moderate number of iterations, e.g., $10^6$, and comparing its final value. For chaotic orbits, SALI rapidly reaches machine precision due to exponential decay, whereas for regular motion, it decreases much more slowly following an algebraic law. For $10^6$ iterations, we expect SALI for regular orbits to remain above approximately $10^{-12}$. Therefore, values below this threshold can be safely interpreted as indicating chaotic dynamics.
    
    Figure \ref{fig:escape_time}(i)--\ref{fig:escape_time}(p) shows SALI computed over the same grid of initial conditions used in Figs.~\ref{fig:escape_time}(a)--\ref{fig:escape_time}(h). Black regions correspond to SALI values below double-precision machine accuracy ($10^{-16}$), indicating chaotic dynamics, while all other colors (from light blue to gray) identify regular motion. A key observation is that many trajectories inside the islands that appear non-escaping in Figs.~\ref{fig:escape_time}(a) and \ref{fig:escape_time}(b) are in fact chaotic. This is clearly revealed by the extended black regions within the islands in Figs.~\ref{fig:escape_time}(i) and \ref{fig:escape_time}(j). These orbits remain trapped for more than $10^8$ iterations, not because they are regular, but due to strong stickiness near remnant invariant structures. Therefore, escape time alone would incorrectly classify these trajectories as regular.
    
    \begin{figure}[t]
        \centering
        \includegraphics[width=\linewidth]{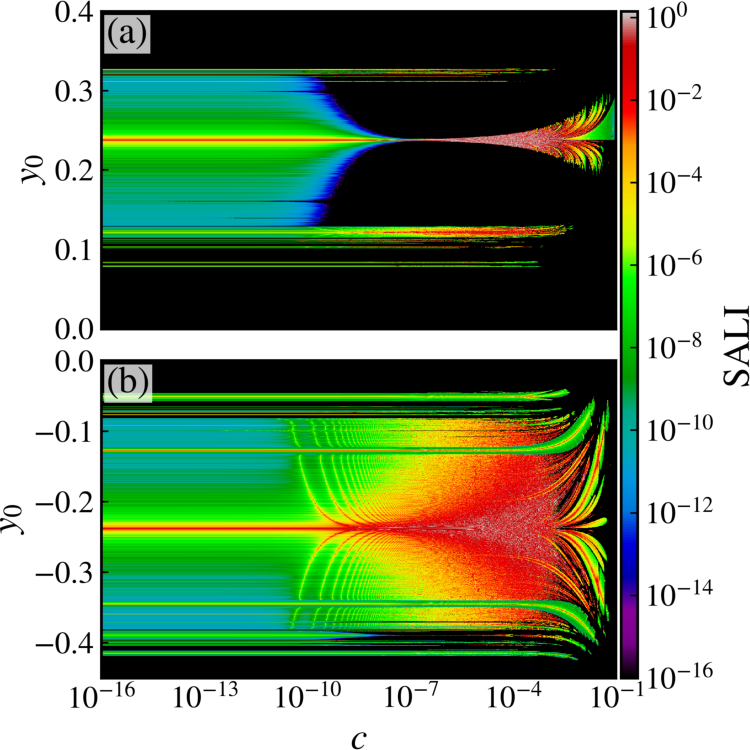}
        \caption{Conservative generalized bifurcation diagrams (CGBDs)~\cite{Manchein2013} illustrating the effect of the perturbation parameter $c$ on the central islands for $x_0 = 0.5$ and $m = 0.8$. Panels (a) and (b) correspond to one-dimensional scans of initial conditions $y_0$ along vertical lines intersecting the upper and lower resonant islands, respectively. For each pair $(c, y_0)$, the smaller alignment index (SALI) was computed up to $10^6$ iterations. The color scale represents the final SALI value, revealing the transition from regular motion (non-vanishing SALI) to chaotic behavior ($\mathrm{SALI} \to 0$) as the perturbation strength $c$ increases.}
        \label{fig:cgbd}
    \end{figure}
    
    Another notable feature is the pronounced asymmetry between the upper and lower island chains. The lower island chain consistently retains significantly larger regular regions, whereas the upper island chain is dominated by chaotic motion even for small values of $c$. Since the two structures correspond to distinct period-$2$ resonances surrounded by different invariant structures, there is no reason to expect that the second perturbation should affect them identically. The observed asymmetry therefore reflects the local nature of the breakup process, with different resonant structures responding differently to the discontinuity of the map on the cylinder. SALI therefore provides a more reliable characterization of the islands than escape time alone, allowing us to quantify how the perturbation parameter $c$ alters the balance between regular and chaotic motion. To investigate this effect systematically, we compute conservative generalized bifurcation diagrams (CGBDs)~\cite{Manchein2013} for the upper [Fig.~\ref{fig:cgbd}(a)] and lower [Fig.~\ref{fig:cgbd}(b)] central islands.
    
    CGBDs are constructed by fixing $x_0 = 0.5$, then computing the SALI up to $10^6$ iterations for (upper chain) $y_0 \in [0.0, 0.4]$, (lower chain) $y_0 \in [-0.42, 0.0]$, and $c \in [10^{-16}, 10^{-1}]$. For extremely small perturbations, $c \in [10^{-16}, 10^{-10}]$, the upper chain remains essentially unaffected. However, for moderately small values of $c$, starting at $c \approx 10^{-10}$, most of the regular region is rapidly replaced by chaotic motion, leaving only a small surviving regular region around the elliptic point. As $c$ increases further, this central regular region expands slightly, but it never recovers the extent observed in the unperturbed case ($c = 0$). In contrast, the lower chain exhibits much greater resistance to the second perturbation. Its regular structure remains largely intact over several orders of magnitude in $c$, and significant destruction of regular regions occurs only for comparatively large perturbations, around $c \approx 10^{-3}$. These results demonstrate that different resonant island chains can respond very differently to the discontinuity of the map on the cylinder. This behavior, however, is not universal: depending on the parameters, the opposite situation can also occur, with the upper island chain being more resistant than the lower one (not shown).

    \begin{figure}[t]
        \centering
        \includegraphics[width=\linewidth]{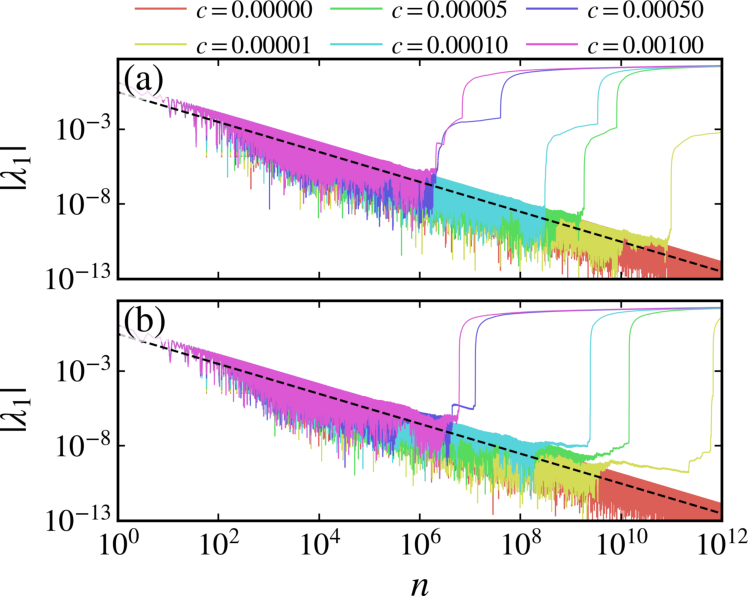}
        \caption{The absolute value of the largest Lyapunov exponent $\lambda_1$ for different values of $c$, computed from an initial condition located at the elliptic point of the central islands for $c = 0$: (a) upper island and (b) lower island. When $c > 0$, the elliptic point may shift or vanish; therefore, the initial condition is kept fixed at the phase-space location of the elliptic point at $c = 0$. The black dashed line corresponds to the slope of $n^{-1}$.}
        \label{fig:mle}
    \end{figure}
    
    The progressive destruction of regular motion under perturbations in Hamiltonian systems is classically described by the KAM and Poincaré-Birkhoff theorems~\cite{lichtenberg2013regular}. Even in smooth, area-preserving maps on the cylinder, this scenario is highly non-trivial: increasing the perturbation typically breaks invariant tori, creates resonant island chains, and enhances transport through the growing chaotic sea. For maps that are continuous on the cylinder, a common picture is that, as the perturbation becomes stronger, more and more KAM curves are destroyed, and the effective size of an island shrinks. Nevertheless, as long as invariant curves persist, the interior of the island remains foliated by invariant curves that act as absolute transport barriers. The island is then bounded by an outermost invariant curve (often referred to as the last KAM curve). When this outer barrier finally breaks, the island does not disappear instantaneously; rather, the boundary retreats inward as a surviving inner invariant curve becomes the new outermost barrier.

    \begin{figure*}[t]
        \centering
        \includegraphics[width=\linewidth]{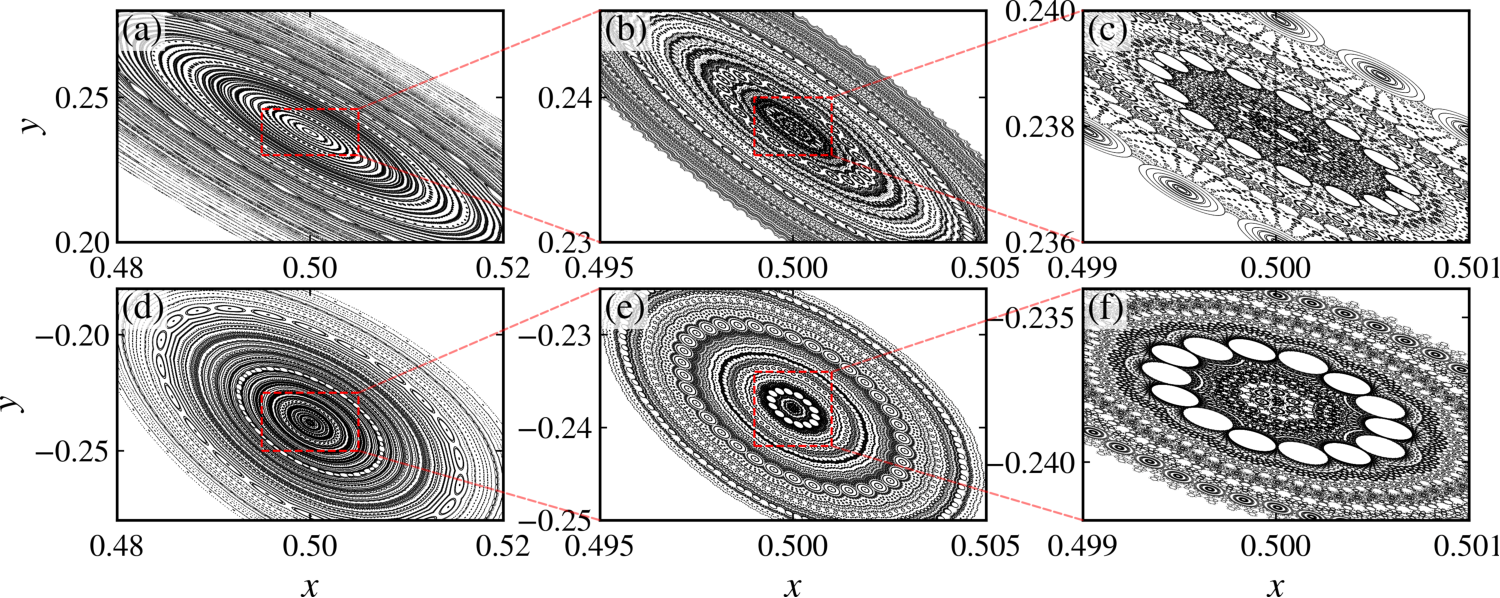}
        \caption{Magnifications around the centers of the upper (left column) and lower (right column) islands for $a = b = 0.53$, $c = 1.0\times10^{-4}$, and $m = 0.8$. Panel (c) shows a magnification of the red dashed box in panel (a), while panel (d) shows a magnification of the red dashed box in panel (b).}
        \label{fig:zoom}
    \end{figure*}
    
    In the present system, the situation is complicated by the discontinuity introduced by non-integer $m$. This raises a central question about the topology of the regular region and the nature of its boundary: does there exist a single global ``last KAM curve'' enclosing the whole island and acting as a complete barrier to transport? Or, instead, does the perturbation destroy the island in a fragmented way, leaving behind only smaller and smaller regular components that are no longer enclosed by a unique invariant curve? To address this question, we probe the innermost region of the islands using initial conditions located at the elliptic fixed points of the unperturbed case $c = 0$. For $c = 0$, the map admits two period-2 elliptic points located at the centers of the upper ($+$) and lower ($-$) central islands:
    \begin{equation}
        \label{eq:upperlowerp2}
        (x_{\pm}, y_{\pm}) = \left(0.5, \pm\sqrt{1-\frac{1}{2a}}\right).
    \end{equation}
    These points are kept fixed as initial conditions while $c$ is varied, and the largest Lyapunov exponent (LLE)~\cite{Benettin1980, Wolf1985, Eckmann1985}, $\lambda_1$, is computed as a function of time for different values of $c$ (Fig.~\ref{fig:mle}). For $c > 0$, the elliptic periodic points generally shift in phase space and may change stability. Therefore, the trajectories are no longer initialized exactly on the periodic orbit, but instead inside the innermost region of the fragmented structure. Our goal is to test whether trajectories initialized deep inside the island remain permanently confined or eventually escape through the fragmented phase-space structure. For $c = 0$, the LLE decays algebraically as $\lambda_1 \propto n^{-1}$, which is the expected decay for the LLE of periodic orbits~\cite{Manchein2013}. When $c > 0$, the LLE initially follows the same algebraic decay, indicating that the trajectory remains temporarily trapped close to a regular region (the stickiness effect). However, after a finite time, $\lambda_1$ exhibits an abrupt increase and converges to a positive value, signaling escape to a chaotic region. This transient algebraic decay followed by a sudden growth of the LLE is characteristic of weakly chaotic dynamics~\cite{Szezech05}. In the case of the upper island [Fig.~\ref{fig:mle}(a)], two distinct abrupt changes in the LLE are observed. The first marks the escape from the innermost regular region into a surrounding chaotic layer, which can be seen in Figs.~\ref{fig:escape_time}(i)--\ref{fig:escape_time}(l), where the LLE stabilizes temporarily at values slightly above $10^{-4}$. The second increase corresponds to a complete escape from the island region altogether.

    Since the initial conditions are chosen deep inside the innermost region of the islands, close to the elliptic periodic points of the unperturbed system, the subsequent abrupt increases and eventual convergence of the LLE to positive values provide strong evidence against the existence of a global transport barrier. Such behavior is incompatible with a globally enclosing invariant curve. If a last KAM curve were present, trajectories initialized near the center of the islands would remain permanently confined, and the LLE would continue to decay algebraically for all times. Therefore, no single ``last KAM curve'' encloses the islands. Instead, the regular regions break up in a fragmented manner, with transport occurring through chaotic channels connecting different spatial scales within the island.
    
    Figure~\ref{fig:zoom} shows successive magnifications of the phase space in the vicinity of the centers of the upper (top row) and lower (bottom row) islands for $c = 10^{-4}$ and $m = 0.8$. At coarse resolution [Figs~\ref{fig:zoom}(a) and \ref{fig:zoom}(d)], the interior of the islands appears qualitatively similar to the continuous case on the cylinder: trajectories lie on apparently closed curves, suggesting quasi-periodic motion organized around elliptic periodic points, with inner resonances in between quasi-periodic trajectories. However, this apparent regularity is misleading. Upon increasing the resolution [Figs~\ref{fig:zoom}(b) and \ref{fig:zoom}(e)], the closed curves are revealed not to be smooth invariant tori, but rather chains of small secondary islands separated by narrow gaps. Further magnification [Figs~\ref{fig:zoom}(c) and \ref{fig:zoom}(f)] shows that these structures form a web-like pattern composed of chains of resonant islands interspersed with chaotic channels through which the trajectories can escape, rather than continuous invariant curves. This hierarchical island chain structure is consistent with a Poincaré-Birkhoff-type breakup of invariant curves and provides direct phase-space evidence of the aforementioned fragmentation within the islands. Similar patterns have been observed in a tilted-hat mushroom billiard~\cite{DaCosta2020} and in a two-degrees-of-freedom Hamiltonian model with two periodic potentials, where the authors named these structures as \emph{myriads}~\cite{Lazarotto2024}.
    
    \section{Finite-time recurrence time entropy analysis}
    \label{sec:ftrte}

    \begin{figure*}[t]
        \centering
        \includegraphics[width=\linewidth]{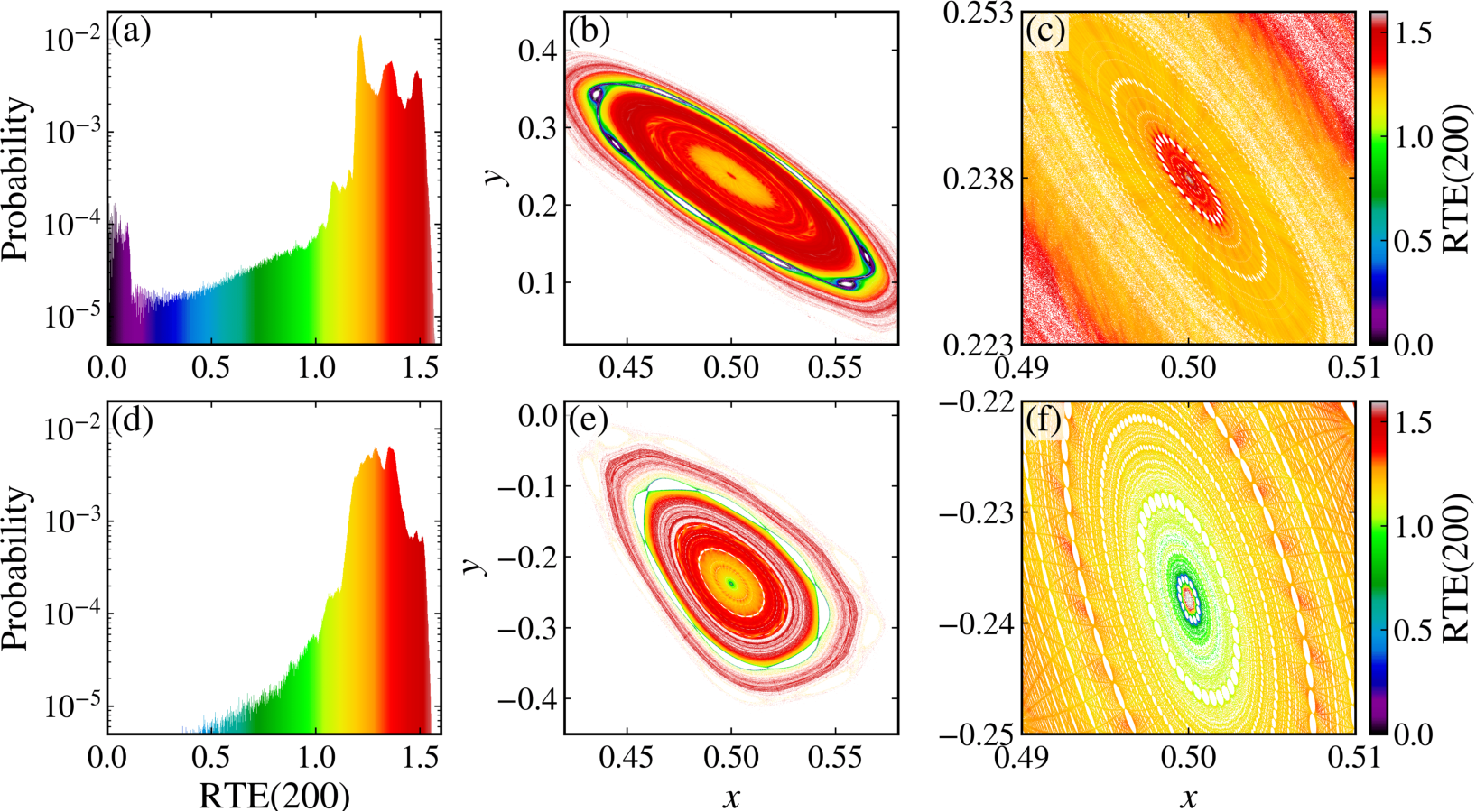}
        \caption{(a) and (d) Finite-time RTE distribution with window length $n = 200$ for an orbit initialized at the elliptic point of the central islands at $c = 0$ (see Fig.~\ref{fig:mle}), iterated until the trajectory escapes to $y = \pm 1$ (see Fig.~\ref{fig:mle} for a time reference), with $a = b = 0.53$, $c = 10^{-4}$, and $m=0.8$. (b) and (e) Phase-space locations of the initial condition for each consecutive window of 200 iterations, colored according to the corresponding $\mathrm{RTE}(200)$ value. (c) and (f) Magnifications around the centers of the islands. The top (bottom) row corresponds to the upper (lower) central island.}
        \label{fig:ftrte}
    \end{figure*}
    
    The phase space magnifications in Fig.~\ref{fig:zoom} reveal that the interior of the regular islands is populated by a hierarchy of small resonant structures separated by narrow chaotic channels. In this fragmented setting, the dynamics is intermittent: a single trajectory alternates between prolonged trapping near island remnants and rapid excursions through chaotic regions, until it eventually escapes to the chaotic sea. Consequently, asymptotic indicators, such as the LLE, are not sufficient to characterize the local dynamical behavior inside the islands. To analyze this intermittent dynamics, we therefore introduce the recurrence time entropy (RTE), obtained from recurrence plots (RPs)~\cite{Kraemer2018}. The RTE was originally introduced independently of RPs~\cite{Little2007} and provides an estimate of the Kolmogorov-Sinai entropy~\cite{Baptista2010}. It is a particularly effective indicator of chaotic behavior because it benefits directly from Slater's theorem~\cite{Slater1950, Slater1967}: quasi-periodic orbits lying on invariant circles with rotation number $\omega$ can have at most three distinct return times, corresponding to the time required for a trajectory to return to the neighborhood of a previously visited state. Additionally, the third return time is always the sum of the other two, and two of them are consecutive denominators in the continued fraction expansion of the irrational rotation number $\omega$.

    There are two standard ways to obtain the recurrence (or return) times of a trajectory. The first consists of selecting an initial condition, defining a neighborhood of size $\varepsilon$ around it, and measuring the time required for the trajectory to return to this region. By iterating the system for a sufficiently long time, one obtains a distribution of recurrence times, from which the Shannon entropy can be computed. An alternative approach is based on recurrence plots (RPs)~\cite{Kraemer2018}, which provide a graphical representation of the recurrences of a $d$-dimensional time series $\mathbf{x}_i = (x^{(1)}_i, x^{(2)}_i, \ldots, x^{(d)}_i)$, where $i = 1, 2, \ldots, N$ and $N$ is the length of the time series. The $N \times N$ recurrence matrix is defined as
    \begin{equation*}
        R_{ij} = H\!\left(\varepsilon - \|\mathbf{x}_i - \mathbf{x}_j\|\right),
    \end{equation*}
    where $i, j = 1, 2, \ldots, N$, $H(\cdot)$ denotes the Heaviside unit step function, $\varepsilon$ is a small threshold, and $\|\mathbf{x}_i - \mathbf{x}_j\|$ is the distance between the $i$th and $j$th states in phase space, computed using a suitable norm. Throughout this paper, we use the maximum $L_\infty$ norm.
    
    The recurrence matrix is a binary matrix that takes the value 1 when the states $i$ and $j$ are closer than $\varepsilon$, indicating a recurrence, and the value 0 otherwise. Several strategies exist for choosing the threshold $\varepsilon$, each with its own advantages and limitations. The two most common approaches are to set $\varepsilon$ as a fraction of the standard deviation of the time series or to select $\varepsilon$ such that a fixed recurrence rate, RR, is obtained. Regardless of the chosen criterion, the influence of a finite $\varepsilon$ cannot be completely eliminated. In this work, we choose $\varepsilon$ so that the recurrence rate is fixed at $\mathrm{RR} = 0.05$.

    The most prominent structures in an RP are diagonal, vertical, and white vertical lines, and a variety of measures based on these structures have been proposed in the literature. For a detailed discussion, we refer the reader to Refs.~\cite{Marwan2002, Marwan2002b, Marwan2007, Marwan2008, Marwan2025} and references therein. Of particular relevance for the present analysis are the white vertical lines, defined as vertical gaps between consecutive diagonal lines. These structures provide a lower bound for the recurrence times~\cite{Zou2007, Zou2007b, Baptista2010, Ngamga2012}. As a result, instead of following a trajectory for very long iteration times, one can analyze a shorter time series, construct its RP, and extract the distribution $P(\ell)$ of white vertical line lengths $\ell$. The entropy of the normalized distribution defines the recurrence time entropy,
    \begin{equation}
        \mathrm{RTE} = -\sum_{\ell = \ell_{\mathrm{min}}}^{\ell_{\mathrm{max}}} p(\ell)\log p(\ell),
    \end{equation}
    where $p(\ell)$ denotes the normalized distribution, and $\ell_{\mathrm{min}}$ and $\ell_{\mathrm{max}}$ are the lengths of the shortest and longest white vertical lines, respectively. Typically, $\ell_{\mathrm{min}} = 1$, while $\ell_{\mathrm{max}}$ corresponds to the longest line found in the RP. For a periodic orbit, there is only a single recurrence time corresponding to the period, which yields $\mathrm{RTE} = 0$. A quasi-periodic orbit admits three recurrence times and therefore produces a small RTE value. In contrast, a chaotic orbit can exhibit a large number of recurrence times, leading to a comparatively large RTE. Trapped orbits fall between these two limiting cases: their RTE values are larger than those associated with quasi-periodic motion, but smaller than those characteristic of fully chaotic trajectories.    
    
    The RTE has been successfully applied to a wide variety of dynamical systems. It has been shown to characterize the dynamics of two-dimensional area-preserving maps~\cite{Sales2023}, fractional maps~\cite{Borin2025}, and both traditional and fractional continuous-time systems~\cite{Gabrick2023}. In particular, it has proven effective in detecting and quantifying stickiness effects and trapping events occurring at different hierarchical levels of the islands-around-islands structure in two-dimensional, area-preserving maps~\cite{Sales2023, Souza2024, Viana2025}. These studies demonstrate that the RTE is especially well-suited for systems in which trajectories exhibit intermittent behavior due to repeated trapping near remnants of invariant structures. Motivated by these results and by the fragmented phase space observed here, we compute the RTE using a finite-time (sliding window) approach, which allows us to resolve temporal variations in the recurrence statistics as the trajectory wanders among small islands and chaotic channels. We choose as our initial conditions the elliptic points for the $c = 0$ case [Eq.~\eqref{eq:upperlowerp2}] and calculate the RTE in windows of size $n = 200$ until each trajectory escapes to $y = \pm 1$. The choice of $n = 200$ follows the standard finite-time recurrence analysis procedure discussed in Refs.~\cite{Sales2023, Souza2024, Viana2025}, representing a compromise between temporal resolution and statistical reliability. Smaller windows lead to noisier recurrence-time distributions, while much larger windows suppress the local variations associated with intermittent trapping and escape. This procedure yields a set of RTE values, from which we construct the corresponding probability distributions, shown in Figs.~\ref{fig:ftrte}(a) and \ref{fig:ftrte}(d). The distributions are broad and strongly asymmetric, reflecting the fragmented phase space. Rather than remaining in a single dynamical regime, the trajectories alternate between long trapping episodes near small stability islands (weakly chaotic regime) and more irregular motion in chaotic regions (stronger chaotic regime). Transitions between these regimes occur through chaotic channels, which are clearly visible in Fig.~\ref{fig:zoom}.
    
    Figures \ref{fig:ftrte}(b) and \ref{fig:ftrte}(e) display the phase space positions of the trajectory at the beginning of each 200 iteration window, with each point colored according to its corresponding RTE value. Low RTE values are concentrated near the centers of the islands and around larger stability regions, while higher RTE values are associated with chaotic channels and with chaotic layers separating chains of islands. This spatial organization shows that the RTE variations are directly linked to the local structures of the fragmented phase space, rather than arising from finite-time estimation effects. The magnifications around the island centers in Figs. \ref{fig:ftrte}(c) and \ref{fig:ftrte}(f) further demonstrate that the RTE varies on very fine spatial scales. Even in the immediate vicinity of the elliptic points, the RTE is not uniform, indicating the presence of internal substructures and partial transport barriers associated with higher-order resonances. This behavior is consistent with the fragmented phase space observed in the magnifications and confirms the highly heterogeneous nature of the dynamics inside the islands, with alternating regions of stronger and weaker chaotic behavior.
    
    \section{Geometrical origin of the fragmentation}
    \label{sec:why}

    \begin{figure*}[t!]
        \centering
        \includegraphics[width=\linewidth]{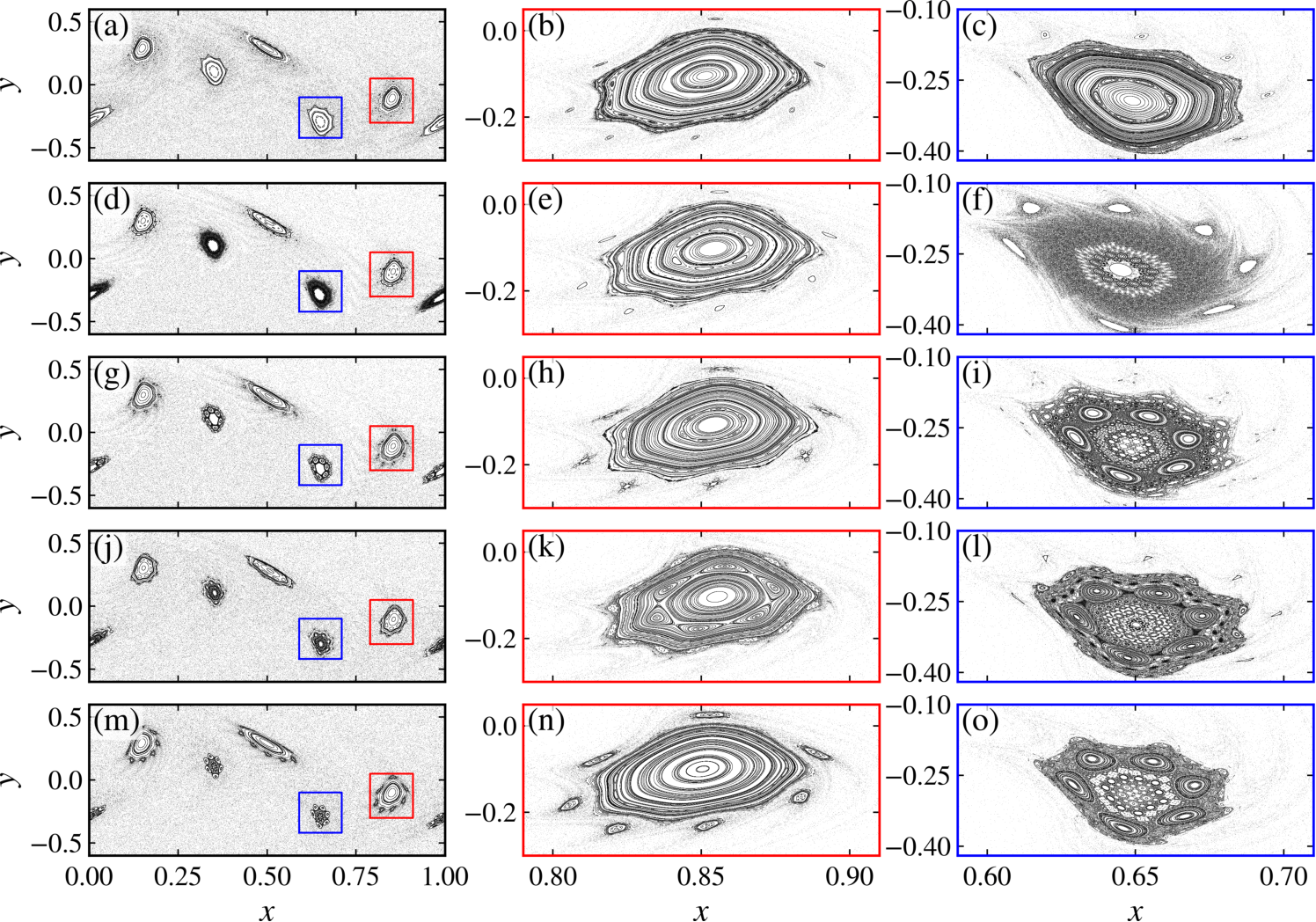}
        \caption{Phase space of the extended standard nontwist map [Eq.~\eqref{eq:esnm}] for $a = 0.71$, $b = 0.494$, $c = 0.005$ and (a)--(c) $m = 0$, (d)--(f) $m = 0.2$, (g)--(i) $m = (\sqrt(5) - 1) / 2$, (j)--(l) $m = 2.6$, and (m)--(o) $m = 3.8$. The middle column corresponds to magnifications of the region inside the red rectangle shown in the left column, while the right column corresponds to magnifications of the region inside the blue rectangle shown in the left column.}
        \label{fig:why}
    \end{figure*}

    The results presented in the previous sections show that the destruction of regular islands in the ESNM does not occur uniformly throughout phase space: different island chains may exhibit qualitatively distinct behaviors under the same perturbation. This raises the question of which geometrical property determines whether a given island chain fragments.
    
    To address this question, Fig.~\ref{fig:why} shows the phase space of the ESNM for parameter values ($a = 0.71$, $b = 0.494$, and $c = 0.005$) in which one of the island chains intersects the discontinuity line at $x = 0, 1$, while another chain remains away from it. The left column displays the phase space around the two chains of period-3 islands, while the middle and right columns show successive magnifications of the regions enclosed by the red and blue rectangles, respectively.
    
    For $m=0$ [Figs.~\ref{fig:why}(a)--\ref{fig:why}(c)], the map is continuous on the cylinder and both island chains exhibit the usual structure of smooth invariant curves surrounding elliptic periodic points. For noninteger values of $m$ [Figs.~\ref{fig:why}(d)--\ref{fig:why}(o)], however, the two island chains behave very differently. The island chain shown in the middle column remains qualitatively similar to the continuous case, with regular trajectories organized around smooth invariant curves. In contrast, the island chain shown in the right column develops the fragmented web-like structure discussed in the previous sections.
    
    The crucial difference between the two cases is their position relative to the discontinuity line. The fragmented island chain possesses one island centered at $x = 0, 1$, whereas the non-fragmented chain remains entirely away from the discontinuity. Since the map is defined modulo $1$, the points $x=0$ and $x=1$ correspond to the same location on the cylinder, forming a discontinuity line for noninteger values of $m$. Island chains intersecting this line are therefore repeatedly cut by the discontinuity introduced by the noninteger perturbation, which prevents the formation of globally enclosing invariant curves and gives rise to the fragmented phase-space structure observed throughout this work.
    
    These results indicate that fragmentation is not a generic consequence of noninteger values of $m$ alone. Instead, the phenomenon depends on the geometrical interaction between the island chain and the discontinuity line. Island chains that do not intersect the discontinuity may remain qualitatively similar to those of continuous area-preserving maps, while chains centered on the discontinuity line exhibit fragmentation and chaotic transport channels at arbitrarily fine scales.

    To further assess the generality of this mechanism, additional phase space plots for different parameter values and distinct area-preserving maps are presented in the Supplementary Material. We consider the extended standard map~\cite{chirikovstdmap, Mugnaine2025, Mugnaine2025b} and the extended Harper map~\cite{Leboeuf1988} with the same type of discontinuity introduced by noninteger values of $m$. In all cases, fragmented structures are observed only for island chains intersecting the discontinuity line at $x = 0, 1$, while island chains located away from this line retain the usual organization around smooth invariant curves. These additional examples indicate that the fragmentation mechanism reported here is not restricted to the ESNM or to nontwist dynamics, but rather constitutes a general geometrical consequence of the interaction between periodic island chains and discontinuity lines in two-dimensional area-preserving maps.
    
    \section{The ESNM on a larger cylinder}
    \label{sec:largercylinder}
    
    \begin{figure*}[t!]
        \centering
        \includegraphics[width=\linewidth]{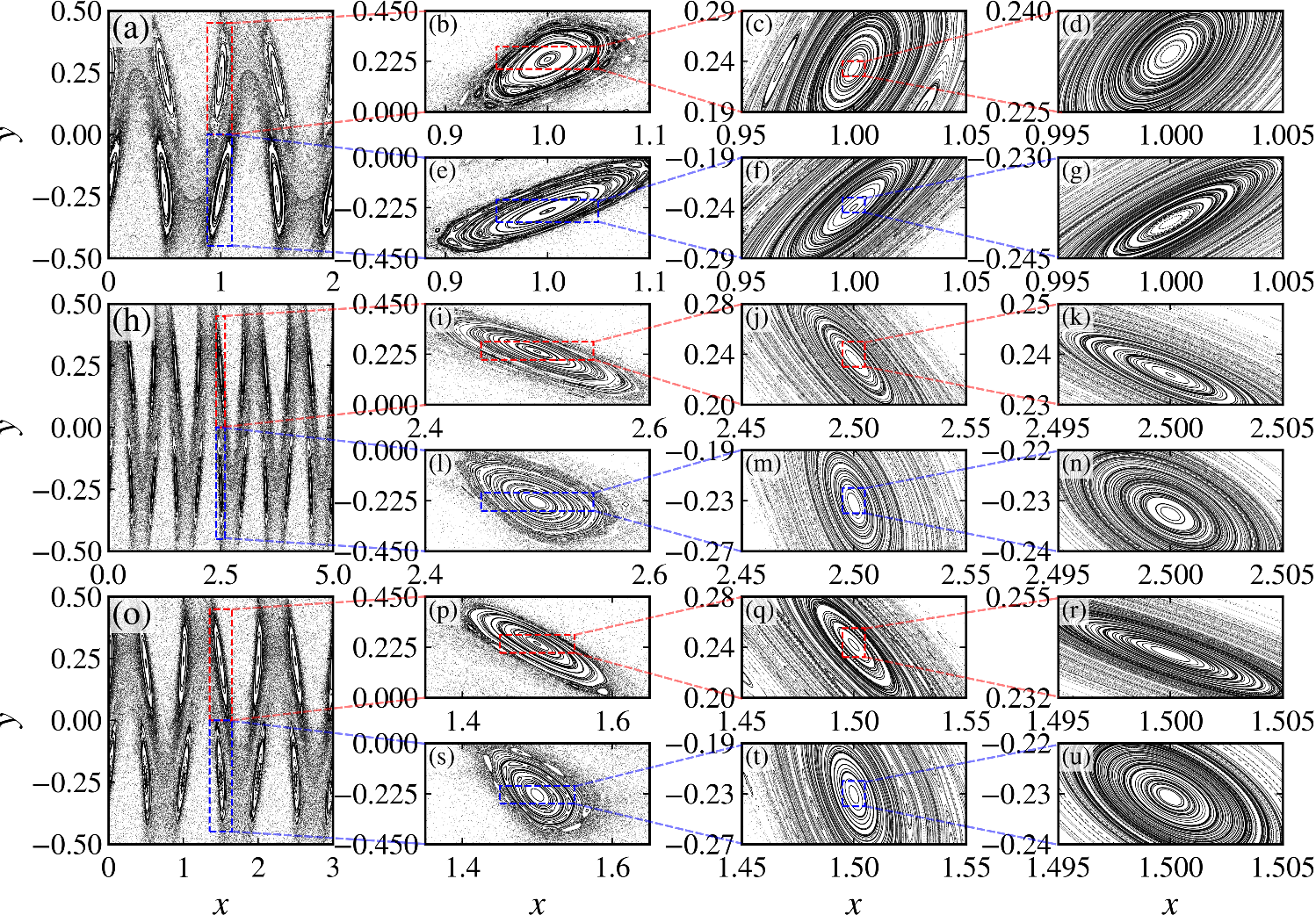}
        \caption{Phase space and successive magnifications inside the stability islands of the extended standard nontwist map defined on a larger cylinder [Eq.~\eqref{eq:modq}] for $a = b = 0.53$, $c = 0.005$, and $m = p / q$, where $p$ and $q$ are integer numbers: (a)--(g) $m = 1/2$, (h)--(n) $m = 4/5$, and (o)--(u) $m = 5/3$.}
        \label{fig:modq}
    \end{figure*}
    
    The discontinuity discussed in the previous sections originates from the term involving the noninteger parameter $m$, which breaks the $1$-periodicity of the map in the angular variable $x$. When $m$ is rational, $m = p/q$ with $p$ and $q$ coprime integers, it is possible to define an alternative formulation in which the map is continuous on the cylinder by enlarging the angular domain. In this case, the map is defined on a larger cylinder, with $x \in [0,q)$ taken modulo $q$, rather than modulo $1$.
    
    With this motivation, we introduce the extended standard nontwist map defined on the cylinder $\mathbb{T}_q \times \mathbb{R}$, where $\mathbb{T}_q = \mathbb{R}/(q\mathbb{Z})$. The map is given by
    \begin{equation}
        \label{eq:modq}
        \begin{aligned}
            y_{n + 1} &= y_n - b\sin(2\pi x_n) - c\sin\!\left( \frac{2\pi p x_n}{q} \right),\\
            x_{n + 1} &= x_n + a\left(1 - y_{n+1}^2\right) \pmod{q}.
        \end{aligned}
    \end{equation}
    In this formulation, the second perturbation is periodic on the interval $[0,q)$, and the induced map on $\mathbb{T}_q \times \mathbb{R}$ is continuous for rational values of $m = p / q$. However, this enlarged cylinder formulation does not preserve the dynamics of the original map defined modulo $1$. Instead, it defines a distinct dynamical system, with different global topology and different invariant structures, even though both systems share the same local functional form.
    
    The purpose of introducing the ESNM on the larger cylinder is therefore not to recover the original dynamics, but to provide a continuous nontwist reference system for rational $m$. This allows us to isolate the effects introduced specifically by the discontinuity in the original formulation by comparing its dynamics with those of a continuous system defined by the same parameters on an enlarged phase space. Figure \ref{fig:modq} shows the phase space of the extended standard nontwist map defined on the larger cylinder [Eq.~\eqref{eq:modq}] for several rational values of $m = p/q$. In contrast with the formulation that is discontinuous on the cylinder discussed in the previous sections, the phase space now exhibits well-organized island chains with smooth boundaries and regular internal structure. The islands appear as elongated, coherent objects that repeat periodically along the extended angular direction, reflecting the enlarged domain $x \bmod q$.
    
    Most importantly, the unusual fragmentation observed in the discontinuous case is no longer present. The successive magnifications of the islands show that the trajectories organize around smooth invariant curves, and the internal structure does not display the web-like pattern or the dense proliferation of gaps that characterized the fragmented phase space. Instead, the islands exhibit the conventional islands-around-islands hierarchy typical of continuous area-preserving maps. The different panels illustrate that, although the number and arrangement of island chains depend on the rational value of $m$, their qualitative structure remains regular across all cases shown. Increasing the denominator $q$ leads to a replication of island chains along the angular direction, but does not induce internal fragmentation. This behavior is consistent with the restoration of continuity on the cylinder and indicates that the fragmentation observed previously is not an intrinsic feature of nontwist dynamics alone.
    
    These observations support the interpretation that the hierarchical fragmentation of islands reported in the original formulation arises from the interaction between the discontinuity line and island chains centered on it, rather than from the nontwist character of the map itself. When the same functional form is embedded in a continuous dynamical system on a larger cylinder, the regular islands persist as coherent structures, and the phase space resembles that of conventional continuous, two-dimensional, area-preserving maps.

    \section{Conclusions}
    \label{sec:concl}
    
    In this paper, we have investigated the dynamical consequences of the discontinuity on the cylinder in the extended standard nontwist map (ESNM) induced by noninteger values of the parameter $m$. We have shown that, when the discontinuity line intersects a resonant island chain, the surrounding invariant structures no longer organize into a sequence of smooth invariant curves. Instead, the phase space develops fragmented structures composed of secondary islands, chaotic layers, and narrow transport channels distributed across multiple spatial scales. This behavior differs qualitatively from the conventional islands-around-islands scenario observed in smooth area-preserving maps defined continuously on the cylinder. Additional examples presented in the Supplementary Material for the standard map and the Harper map show that the same qualitative mechanism also appears in other discontinuous area-preserving systems. Similar structures have also been reported in a tilted mushroom billiard~\cite{DaCosta2020} and in a two-degrees-of-freedom Hamiltonian system with two periodic potentials~\cite{Lazarotto2024}. The phase space of these systems likewise exhibits web-like structures and persistent partial transport barriers for certain parameter values. Whether these phenomena share a common underlying mechanism remains an open question.

    By calculating the escape times, we have shown that many trajectories remain trapped for extremely long times, up to $10^8$ iterations, which would normally suggest regular motion. By computing the smaller alignment index (SALI), we have demonstrated that a large fraction of these nonescaping trajectories are in fact chaotic. Stickiness near remnant invariant structures delays escape, and SALI therefore provides a necessary diagnostic to distinguish true regular motion from weak chaos in this fragmented setting.

    We further quantified the destruction of regular motion using conservative generalized bifurcation diagrams (CGBDs). These revealed a strong asymmetry between the upper and lower period-$2$ island chains. We have shown that this asymmetry is a consequence of the local nature of the breakup process: although both structures correspond to period-$2$ resonances, they are distinct island chains surrounded by different invariant structures, and therefore the discontinuity affects them differently. We identified cases in which one island chain fragments while the other preserves the conventional hierarchical organization observed in continuous maps.

    By computing the largest Lyapunov exponent (LLE) from initial conditions placed at the elliptic points of the continuous system ($c = 0$), we directly tested whether a global transport barrier survives inside the islands. For $c > 0$, the LLE exhibits transient algebraic decay followed by abrupt growth to a positive value, signaling escape from successive trapping regions. This behavior rules out the existence of a single ``last KAM curve'' enclosing the islands. Instead, the breakup occurs locally through the formation of chaotic channels connecting different regions of phase space across multiple scales. To characterize the strongly intermittent dynamics generated by this fragmented phase space, we introduced a finite-time recurrence-time entropy analysis (RTE). We found broad, asymmetric distributions reflecting repeated transitions between weakly chaotic trapping near island remnants and stronger chaotic motion in surrounding layers. Mapping finite-time RTE values onto phase space revealed a clear spatial organization: low RTE values cluster near larger regular structures, while high RTE values are concentrated along chaotic channels and strongly chaotic regions. These variations persist even at fine spatial scales, confirming that the fragmentation extends deep into the hierarchical structure.

    Finally, we introduced a continuous reference system by defining the ESNM on a larger cylinder for rational values of $m = p/q$. Although this formulation does not preserve the original dynamics, it isolates the role of the discontinuity on the cylinder by restoring continuity on the enlarged cylinder while keeping the same functional form. In this continuous system, the fragmentation disappears entirely and the phase space recovers the conventional organization of smooth area-preserving maps, with regular island chains bounded by smooth invariant curves. This comparison demonstrates that the observed fragmentation is a direct consequence of the discontinuity on the cylinder.

    The simulations presented in this paper were performed using the \textit{pynamicalsys} package~\cite{pynamicalsys}, a Python toolkit for the analysis of dynamical systems.

    \section*{Supplementary Material}

    See the Supplementary Material for additional phase space plots of the extended standard nontwist map for different parameter values, as well as for the extended standard map and the extended Harper map. 
    
    \section*{Code availability}
    
    The source code to reproduce the results reported in this paper is available online in the GitHub repository:
    \href{https://github.com/mrolims-publications/fragmentation-esnm}{https://github.com/mrolims-publications/fragmentation-esnm}.

    \section*{Acknowledgments}
    
    This work was supported by the São Paulo Research Foundation (FAPESP, Brazil), under Grant Nos.~2019/14038-6, 2021/09519-5, 2023/08698-9, 2023/16146-6, 2024/05700-5, and 2024/20417-8, by the National Council for Scientific and Technological Development (CNPq, Brazil), under Grant Nos.~302665/2017-0, 301318/2019-0, 309670/2023-3, and 304398/2023-3, and by the Paraná Research Foundation (Fundação Araucária). The computational simulations in this research were supported by resources supplied by the Center for Scientific Computing (NCC/GridUNESP) of the São Paulo State University (UNESP) and by the \href{http://portal.if.usp.br/controle/}{Oscillations Control Group} of the University of São Paulo. We would also like to thank the \href{https://www.105groupscience.com}{105 Group Science} for fruitful discussions.

\end{document}